\documentclass[prl,aps,amssymb,showpacs,twocolumn,superscriptaddress]{revtex4}

\usepackage{graphicx,amssymb,amsmath,color}
\usepackage{amsthm,bm}
\usepackage{amsfonts}

\begin{document}


\title{Unusual hyperfine interaction of Dirac electrons and NMR spectroscopy in graphene}

\author{Bal\'azs D\'ora}
 \affiliation{Max-Planck-Institut f\"ur Physik Komplexer Systeme, N\"othnitzer Str. 38, 01187 Dresden, Germany}
\author{Ferenc Simon}
\affiliation{Budapest University of Technology and Economics, Institute of Physics and Solids in Magnetic Fields
Research Group, Hungarian Academy of Sciences, P.O. Box 91, H-1521 Budapest, Hungary}

\date{\today}

\begin{abstract}
Theory of nuclear magnetic resonance (NMR) in graphene is presented. The canonical form of the electron-nucleus hyperfine interaction is
strongly modified by the linear electronic dispersion. The NMR shift and spin-lattice relaxation time are calculated
as function of temperature, chemical potential, and magnetic field and three distinct regimes are identified: Fermi-,
Dirac-gas,
and extreme quantum limit behaviors. A critical spectrometer assessment shows that NMR is within reach for fully $^{13}$C
enriched graphene of reasonable size.
\end{abstract}

\pacs{76.60.-k,71.20.Tx,85.75.-d}

\maketitle


Nuclear magnetic resonance (NMR) is a powerful spectroscopic tool \cite{SlichterBook} and an
architecture for quantum
information processing \cite{NMR_quantcomp,NMR_quantcomp_exp} at the same time.
Both of these applications are possible due to the relatively weak interaction of the nucleus with its environment.
This weak interaction is sufficient to probe the electronic state of its vicinity,
which yields information about the local electron bonds or about the correlated behavior of electrons as e.g. in
superconductors \cite{HebelSlichter}. NMR quantum computing exploits that the nuclei are well isolated from the
environment thus there is a longer time window for the manipulation and detection of the nuclear quantum state.

For both kinds of applications, the important NMR parameters are the shift of the NMR resonance with respect to a standard, and the decay of
the longitudinal magnetization to its equilibrium value, the spin-lattice relaxation time, $T_1$. These were extensively
studied in solid state systems both theoretically and experimentally \cite{abragam,SlichterBook}. However, the body of
NMR experiments were focused on three-dimensional systems which stemmed from the unavailability
of stable, inherently two-dimensional materials.
The discovery of graphene, a single stable sheet of carbon atoms in a hexagonal lattice \cite{novoselov1},
enables studies of an exactly two-dimensional system.
Its quasi-particles follow a linear band dispersion, causing the electrons to behave as massless Dirac fermions,
which gives rise to unique transport and magnetic properties\cite{rmpguinea}. Similarly, unusual electron-nuclear interaction is expected.


In a metal, the NMR measurables are most affected by the surrounding electrons through the electron-nuclear
hyperfine interaction (HFI). The standard, text-book form of the HFI of nuclei and conduction electrons
leads to the Hamiltonian $H_{\text{HFI}}=H_{\text{orb}}+H_{\text{spin}}$ \cite{abragam}:
\begin{gather}
H_{\text{orb}}=\frac{\mu_0}{4\pi} g \mu_{\text{B}}^* \gamma_n{\bf I}  \frac{{\bf r}\times{\bf p}}{ r^3} ,\nonumber\\
H_{\text{spin}}=\frac{\mu_0}{4\pi} g \mu_{\text{B}} \hbar \gamma_n{\bf I} \left(\frac{{\bf S}r^2-3\bf r(Sr)}{r^5}-\frac{8\pi}{3}{\bf S}\delta(\bf r)\right)
\label{NormalHFI}
\end{gather}

\noindent Here, the first term ($H_{\text{orb}}$) is due to the electron orbital magnetism,
the second ($H_{\text{spin}}$) contains  the electron spin-dipole interaction  and the so-called Fermi-contact interaction. $\mu_0$ is the permeability of
free space, $\gamma_n$ is the nuclear gyromagnetic ratio, $\bf I$ is the nuclear spin, $g\approx 2$ is the g-factor of the electrons. \textbf{S}, $\bf p$, and
$\bf r$ are the electron spin, momentum, and vector operators. $\mu_{\text{B}}$ is the Bohr magneton and $\mu_{\text{B}}^*=m/m^*\mu_{\text{B}}$ is the effective orbital Bohr magneton \cite{lee}, where $m^*$ is the effective band mass and $m$ is the mass of a free electron.

At first, it is not obvious how to generalize the orbital term of the HFI to massless
Dirac fermions 
and its derivation is one of the primary goals of this work.
Second, the unique properties of the conduction electrons of graphene
are expected to give rise to unique relaxation and NMR shift behaviors. E.g. deviation from the
Korringa relation \cite{SlichterBook}, that is an important benchmark of non Fermi-liquid behavior, is expected.

Here, we show that the canonical description of the hyperfine interaction is modified for the massless Dirac fermions and
we derive the hyperfine Hamiltonian paying special attention to obtain the appropriate orbital contribution. We identify different regimes based on the NMR measurables; Fermi-,
Dirac-gas, and extreme quantum limit behaviors. We also discuss the feasibility
of bulk NMR spectroscopy on graphene with a critical evaluation of NMR spectrometer performance.


The low energy excitations in graphene are described by the two-dimensional Dirac equation \cite{rmpguinea}:
\begin{equation}
H=v_{\text{F}}(\sigma_xp_x+\sigma_yp_y),
\label{diracham}
\end{equation}
\noindent where $v_{\text{F}}\approx 10^6$~m/s is the Fermi velocity of graphene, and the pseudospin variables (Pauli matrices, $\sigma$) spring
from the two-sublattice structure. The HFI in graphene is derived following Abragam \cite{abragam} by
treating the nucleus as a magnetic dipole with ${\bf m}=\hbar\gamma_n \bf I$. Its vector potential, ${\bf A(r)}=\frac{\mu_0}{4\pi}\frac{{\bf m}\times \bf r}{r^3}$, is inserted into the kinetic momentum as ${\bf p} \rightarrow {\bf p}+e{\bf A}$ in addition to the electron and nuclear Zeeman terms. This calculation gives the effective HFI Hamiltonian in graphene as

\begin{gather}
{H}_{\text{HFI}}^\text{gr}=\frac{\mu_0}{4\pi}\hbar\gamma_n{I_z}\left(\frac{{\bf r}\times{\bf j}}{r^3}\right)_z+H_{\text{spin}},
\label{heff}
\end{gather}
\noindent where ${\bf j}=ev_{\text{F}}\bm\sigma$ is
the electric current operator in graphene and $\bm\sigma$ is a vector of the Pauli matrices. The first term describes the
interaction of the nucleus with the orbital motion of the Dirac electrons\cite{klink}, which contains $I_z$ only as the electrons are confined in the plane. The spin-dipole and Fermi-contact terms are unchanged with respect to their usual forms.

The orbital term in Eq. \eqref{heff} differs significantly from the usual form in Eq. \eqref{NormalHFI} as the orbital magnetic moment (${\bf
r}\times{\bf j}$) replaces the usual term ($\frac{g \mu^*_{\text{B}}}{\hbar}{\bf r}\times{\bf p}$). This is the result of the peculiar form of the current operator for Dirac electrons $\bf j\sim \bm \sigma$, which is also is responsible for the jittery motion of the center of
mass coordinate known as Zitterbewegung \cite{cserti}.
Eq. \eqref{NormalHFI} can be deduced formally from Eq. \eqref{heff} by using ${\bf j}=e{\bf p}/m^*$ for a normal metal.

A unique property of the orbital magnetic moment of graphene is that it remains invariant in an applied magnetic or gauge field, since $\bf j$ is insensitive to the vector potential. We mention that the
proper orbital angular momentum of Dirac particles is still ${\bf r}\times {\bf p}$ in the sense that it is responsible for rotations
in the $x-y$ plane, which differs from the orbital magnetization. We also note that there are no higher order terms in the vector potential in the graphene HFI Hamiltonian due to the linearity of the Dirac equation.




The second quantized form of the orbital part of the interaction in graphene is obtained as
\begin{gather}
H_{\text{orb}}^\text{gr}=\frac{J_{\text{orb}}}{N}I_z\sum_{{\bf kk}'\alpha\alpha' s}f({\bf k,k'},\alpha,\alpha')
c^+_{{\bf k}\alpha s}c_{{\bf
k'}\alpha's},
\end{gather}
where $J_{\text{orb}}=\mu_0 \hbar\gamma_n ev_{\text{F}}/2 A_c$, and
$f({\bf k,k'},\alpha,\alpha')={(\alpha\alpha'-\exp[i(\varphi_k-\varphi_{k'})])(\alpha k+\alpha'k')}/{2|\bf k-k'|},$
and $c^+_{{\bf k}\alpha s}$ creates a quasi-particle with energy $E_{\alpha}(\bf k)$ and real spin $s$,
$\varphi_k$ is the angle of $\bf k$ with the $k_x$ axis, $A_c$ is the unit cell area,
and $N$ is the number of unit cells.
The interaction is bounded as $|f({\bf k,k'},\alpha,\alpha')|\leq 1$.
The magnitude of the orbital term is estimated as $J_{\text{orb}}\approx 21$~MHz using $\gamma(^{13}\text{C})/2\pi= 10.7$~MHz/T.


The effective interaction describing the hyperfine interaction in graphene is obtained from Eq. \ref{heff}. as
\begin{equation}
H_{\text{HFI}}^\text{gr}={\bf S \bar A I}+H_{\text{orb}}^\text{gr},
\label{HFIgraphene}
\end{equation}
where $\bf \bar A$ is a 3x3 tensor with diagonal elements. Of these, the traceless ones are due to the spin-dipole interaction as
$A_{\text{dip}}(x,y):A_{\text{dip}}(z)=-A_{\text{dip}}:2A_{\text{dip}}$ and the scalar term, $A_{\text{iso}}$, is given by the isotropic Fermi-contact interaction. First principles calculations \cite{yazyev} gave $A_{\text{dip}}=73$ MHz and $A_{\text{iso}}=-44$ MHz, which gives $(-117,-117,102)$~MHz for the diagonal elements of $\bf \bar A$. We note that the first principles value of $A_{\text{dip}}$ agrees well with the $A_{\text{dip}}=91$~MHz obtained for the $p_z$ orbital of a free carbon
atom \cite{SlichterBook,sato}, which confirms that it is indeed the relevant orbital in graphene.

Upon establishing the hyperfine interaction in graphene, we turn to the calculation of the NMR measurables. 
For a given magnetic field, terms of Eq. \ref{HFIgraphene}. perpendicular and parallel to the
field contribute to relaxation and to the Knight shift, respectively \cite{abragam,winter}.
The spin-lattice relaxation rate, $1/T_1$ and the Knight shift, $K$, for a given magnetic field direction ($i=x,y,z$) are \cite{moriya}
\begin{subequations}
\begin{gather}
\left(\frac{1}{T_1T}\right)_i=\frac{C_{i}^2\pi k_{\text{B}}}{\hbar }\int\limits_{-\infty}^\infty
\frac{\rho(E)^2 \textmd{d}E}{4k_{\text{B}}T\cosh^2[(E-\mu)/2k_{\text{B}} T]},\label{t1relax}\\
K_i=\frac{A_i \gamma_e}{2\gamma_n}\int\limits_{-\infty}^\infty
\frac{\rho(E)\textmd{d}E}{4k_{\text{B}}T\cosh^2[(E-\mu)/2k_{\text{B}}T]},
\end{gather}
\end{subequations}
with $C_i^2=\sum_{\nu\neq i}(A_{\nu}^2/2+\delta_{\nu,z}2J_{\text{orb}}^2)$, $\gamma_e$ is the gyromagnetic ratio of
electrons, $\rho(E)$ is the quasi-particle density of states (DOS), and $\mu$ is the chemical potential.
$T_1$ and $K$ for an arbitrary field direction is readily obtained by angular dependent combinations \cite{winter}.

The orbital interaction involves only $I_z$, thus it affects $T_1$ only when the field is in
the graphene plane ($i=x,~y$), which explains the $2J_{\text{orb}}^2$ term (the factor 2 comes from the spin
degeneracy). The orbital term does not contribute to the Knight shift even for a magnetic field along $z$ in a manner analogous to demagnetization. The spin part of the HFI contributes to a $\sim 15$ \% anisotropy of $T_1$ for in and out of plane magnetic fields but the orbital term makes it nearly isotropic. More accurate statements require the first principles calculation \cite{yazyev} of $J_{\text{orb}}$. The Knight shift changes sign and drops by 15 \% from in plane to out of plane fields. We omit the $i$ index from $C_i$ in the following.

We distinguish two scenarios for the DOS in the following calculation: (i) absence of Landau levels and (ii) where the presence of Landau levels is important.
Scenario (i) occurs for three cases: when magnetic field is in the plane, when magnetic field is arbitrary but
level broadening due to $\Gamma$ or $T$ makes the Landau levels undistinguishable around $\mu$ (the criterion is
$v_{\text{F}}^2eB\mu \leq$max$(\Gamma,k_{\text{B}}T)$), or in the vicinity of the DP point (i.e. $\mu$ is small) when
the lowest Landau level is significantly broadened due $\Gamma$ or $T$ (the criterion is
$v_{\text{F}}\sqrt{2eB\hbar}\leq$max$(\Gamma,k_{\text{B}} T)$).

For scenario (i), the magnetic field-free DOS can be used in the calculation and it reads as:

\begin{equation}
\rho(E)=\frac{A_c|E|}{2\pi\hbar^2 v_{\text{F}}^2}
\label{freedos}
\end{equation}
\noindent per spin and C atom.
The resulting relaxation rate is
\begin{equation}
\left(\frac{1}{T_1T}\right)= C^2\frac{\pi k_{\text{B}}}{\hbar}\left[\rho^2(\mu)+\rho^2\left(\frac{\pi k_{\text{B}}T}{\sqrt
3}\right)\right],
\label{relax}
\end{equation}
which increases as max($\mu^2,(\pi k_{\text{B}}T)^2/ 3$).
Away from the Dirac point, $\mu$ dominates, and the temperature
becomes important only near the DP.

Exactly at the DP, $T_1$ diverges as $T_1\sim T^{-3}$,
therefore the nuclear spins are not relaxed by conduction electrons at $T=0$ due to the absence of charge carriers at the
charge neutrality point. 
In the presence of impurities, the DOS at the DP reads as \cite{sharapovdos}
\begin{gather}
\rho(0)=\frac{A_c}{2\pi\hbar^2 v_{\text{F}}^2}
\frac{2\Gamma}{\pi}
\ln\left(\frac{D}{\Gamma}\right)
\end{gather}
\noindent with $\Gamma$ the scattering rate and $D$ the cutoff in the continuum theory.
Therefore, the aforementioned divergence of the clean system weakens to $T_1\sim (\Gamma^2\ln^2(D/\Gamma) T)^{-1}$,
reproducing the Fermi-gas behavior. Since the DOS is finite at the DP due to
impurities, the Dirac nature of the quasi-particles is lost at this level.

The Knight shift is evaluated as
\begin{equation}
K=A\frac{\gamma_e}{2\gamma_n}
\rho
\left(2k_{\text{B}}T \ln\left[2\cosh\left(\frac{\mu}{2k_{\text{B}}T}\right)\right]\right).
\end{equation}
It can be approximated by $K\sim$ max($2k_{\text{B}}T\ln 2,|\mu|$).
Impurities provide a finite DOS even at the DP, therefore the Knight shift stays finite there as $K\sim \Gamma\ln(D/\Gamma)$.

\begin{figure}[t!]


\centering{\includegraphics[width=6cm,height=5cm]{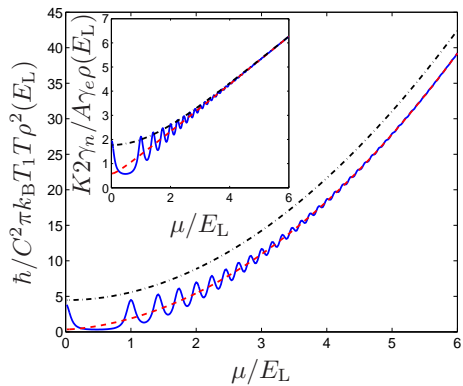}}
\caption{(Color online) The nuclear spin relaxation rate (main figure) and the Knight shift (inset) are shown with $\Gamma=0.1 ~E_{\text{L}}$ and
$D=1000 ~E_{\text{L}}$ as a function of the chemical potential. The blue solid/red dashed line refers to the
presence/absence of magnetic field at $T=0$, the black dash-dotted line corresponds to $k_{\text{B}}T=E_{\text{L}}$ in the presence of magnetic field.
Increasing $\mu$ or $T$ makes the Landau level structure disappear, and scenario (ii) is replaced by scenario (i).
\label{t1kfield}}
\end{figure}

Concluding scenario (i), we give $1/T_1T$ for the case of chemical doping of, or chemisorption on the graphene layer. E.g. for an AC$_{x}$
composition, where A is an alkali atom with full charge transfer, there is an extra $2/xA_c$ electron density to each lattice site.
This translates to a chemical potential shift of $\mu=\hbar v_F\sqrt{2\pi/xA_c}$, which
leads to a relaxation rate as
\begin{equation}
\left(\frac{1}{T_1T}\right)=\frac{k_{\text{B}} C^2 A_c}{2\hbar^3v_{\text{F}}^2x}\approx \frac{0.002}{x}[\text{(Ks)}^{-1}],
\end{equation}
or $T_1\approx 500\text{ (sK)}\cdot x/T$. 
This gives $T_1 \approx 10$ s at 300 K for $x$=8, 
that is a usual doping level for graphite \cite{DresselhausAP2002}. It shows 
the sensitivity of the NMR properties for doping or chemisorption, which may 
lead to a sensor application of graphene. The chemical potential can be also 
tuned by gate voltage with a less dramatic effect on $T_1$.

For scenario (ii), Landau level formation is important, and the continuous spectrum is replaced by discrete Landau levels as
$E_{n\alpha}=\alpha E_{\text{L}}\sqrt{n}$ where $\alpha=\pm$, $n$ is non-negative integer, $E_{\text{L}}=v_{\text{F}}\sqrt{2\hbar eB_z}$ is the Landau
scale, and $B_z$ is the perpendicular component of the magnetic field. With this, the DOS reads
as\cite{sharapovdos}
\begin{gather}
\rho(E)=\frac{A_c}{2\pi\hbar^2
v_{\text{F}}^2}\frac{1}{2\pi}\left[\frac{\Gamma E_{\text{L}}^2}{E^2+\Gamma^2}
-4\Gamma\ln\left(\frac{E_{\text{L}}}{D}\right)-\right.\nonumber\\
\left.-2\textmd{Im}\left\{(E+i\Gamma)\Psi\left(1-\frac{(E+i\Gamma)^2}{E_{\text{L}}^2}\right)\right\}\right],
\label{LandauDOS}
\end{gather}
where $\Psi(x)$ is Euler's digamma function, and reduces to Eq. \eqref{freedos} for clean systems with vanishing magnetic field.
The $1/T_1 T$ and the Knight shift are shown in Fig. \ref{t1kfield} separately as a function of the chemical
potential. These display the characteristic de Haas-van Alphen
like oscillatory behavior at high magnetic field and low $\mu$ and $T$, that we refer to as the extreme quantum
limit (EQL).

\begin{figure}[t!]


\centering{\includegraphics[width=4cm,height=4cm]{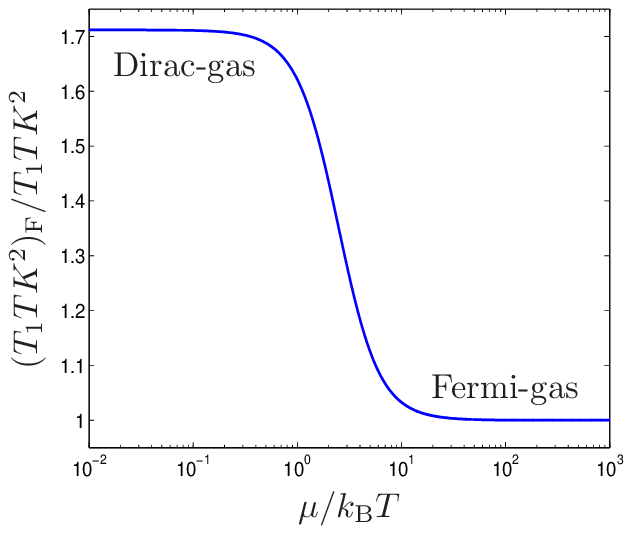} \includegraphics[width=4cm,height=4cm]{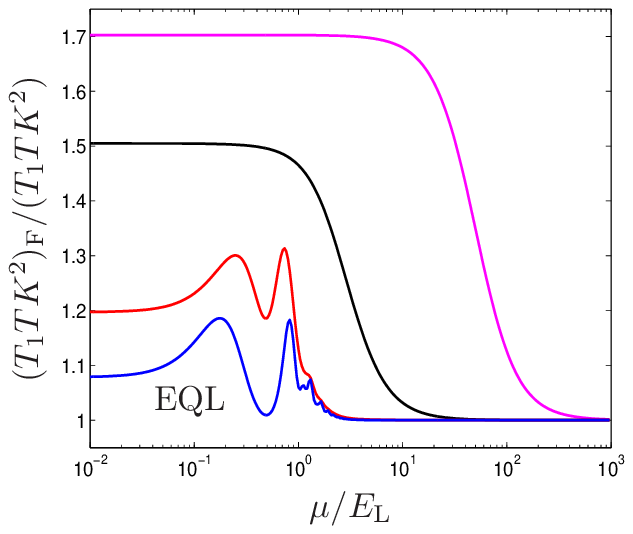}}
\caption{(Color online) The Korringa relation, normalized to its Fermi-gas value, as function of $\mu/k_{\text{B}}T$ (left panel) and $\mu/E_{\text{L}}$ (right panel).
The usual Korringa relation is recovered for $\mu\gg k_{\text{B}}T$ but for increasing $T$ the normalized Korringa relation increases and saturates to $\pi^2/3\ln^2(4)$. The right panel shows the Korringa relation for $\Gamma=0.1 ~E_{\text{L}}$, which allows the visibility of the lowest Landau levels. The temperature is
varied as $k_{\text{B}}T/ E_{\text{L}}$=0.05, 0.1, 1 and 20 from bottom to top.
For $k_{\text{B}}T\sim\Gamma$, separate peaks indicate the Landau level structure, and the curves cross over to the
field-free scaling limit with increasing $T$.
\label{korringalandau}}
\end{figure}

Calculation of the relaxation rate and Knight shift allows to test the validity of the Korringa relation,
i.e. whether $1/T_1TK^2=\text{const}$. holds. In general, the Korringa relation is valid for a Fermi-liquid. In particular for a non-interacting Fermi-gas \endnote{It is shown in Ref. \cite{Antropov} that neither dipolar nor orbital
anisotropy affects the Korringa relation} $(1/T_1TK^2)_{\text{F}}=4\pi k_{\text{B}} (\gamma_n/\gamma_e)^2/\hbar$.
For graphene within scenario (i) and in the limit of $(\mu,k_{\text{B}}T)\gg \Gamma$, which is referred to as the scaling
limit, it reads as
\begin{equation}
\frac{1}{T_1TK^2}=\frac{4\pi k_{\text{B}}}{\hbar}\left(\frac{\gamma_n}{\gamma_e}\right)^2 \left(\frac{C^2}{A^2}\right)F\left(\frac
{\mu}{k_{\text{B}}T}\right),
\end{equation}
which depends only on the ratio of $\mu$ and $T$, and $F(x)$ is a universal scaling function
\begin{equation}
F(x)=\frac{3x^2+\pi^2}{12\ln^2[2\cosh(x/2)]},
\end{equation}
which is even in $x$ and satisfies $F(0)=\pi^2/3\ln^2(4)\approx 1.71$ and $F(\infty)=1$, and is shown in Fig.
\ref{korringalandau}.
For in-plane field, $C^2/A^2<1$, while for perpendicular field $C^2/A^2>1$.
For $\mu\gg k_{\text{B}}T$, the DOS is finite, and nothing distinguishes graphene from a conventional metal because
only one
branch of the "V"-shaped dispersion is seen due to the smallness of $T$, therefore the usual
Korringa relation is satisfied.
In the opposite limit, ($\mu \ll k_{\text{B}}T$), the Korringa relation leads to a constant, $F(0)$, times bigger
than its
conventional value, which signals the nature of Dirac fermions.
The crossover can be explored even away from the DP by fixing the chemical potential to a finite value, and sweeping the
temperature.
Right at the DP, impurities spoil the crossover and re-establish the Fermi-gas relation for $(k_{\text{B}}T,\mu)\ll \Gamma$.

The Korringa relation can be numerically evaluated in the presence of Landau levels (i.e. for scenario (ii)) using Eq. \ref{LandauDOS} and Fig. \ref{korringalandau} shows the result. For small
$T$ and $\Gamma$, the oscillatory behavior due to Landau levels in the DOS characterizes the Korringa
relation. When $(k_{\text{B}} T,\Gamma)>E_{\text{L}}$, the Landau levels are smeared and the magnetic field does not play an important role thus the scaling limit is restored.

\begin{figure}[t!]

\centering{\includegraphics[width=4cm,height=4cm]{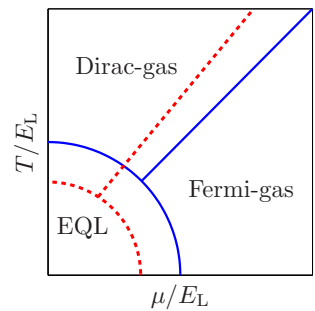}}
\caption{The schematic "phase diagram" of graphene according to NMR, the phases are separated by solid lines.
The boundaries denote smooth crossovers. The solid lines move to the dashed ones for increasing disorder.
\label{phase}}
\end{figure}

In Fig. \ref{phase}, we summarize our findings on the NMR properties in the form of a "phase diagram". The extreme quantum limit shows up only at low temperatures and
small chemical potential, when the Landau level structure is visible. Larger $\Gamma$, i.e. presence of defects, favors the Fermi-gas region.

We finally comment on the feasibility of NMR experiments in graphene. NMR is known to have a low signal sensitivity albeit its
tremendous utility. In graphene, the NMR active $^{13}$C nuclei has a low abundance ($c=1.1~\%$) and a low gyromagnetic
ratio, $\gamma(^{13}\text{C})\approx \gamma(^1\text{H})/4$. NMR spectrometers are characterized by the limit
of detection (LOD) parameter, i.e. the number of nuclei required for a signal-to-noise ratio of three
in a single acquisition. State-of-the-art spectrometers \cite{vanBentumJMR2007} have $\text{LOD}_0=10^{12}/\sqrt{\text{Hz}}$
for $^{1}\text{H}$ spins with sample and detector at 300 K in a 14 T magnetic field ($\nu(^1\text{H})=600$ MHz).
For a general case the LOD is \cite{MassinJMR2003}:

\begin{equation}
\text{LOD}=\frac{\text{LOD}_0 }{c} \sqrt{\frac{1\text{ sec}}{T_2^*}}\left( \frac{\gamma(^1\text{H})}{\gamma}\right)^3\frac{T_s}{300 \text{ K}} \text{NF}_{\text{rel}}
\label{NMRsensitivity}
\end{equation}

\noindent Here $T_2^*$ is the apparent decay time of the NMR time-domain signal which
contains the spin-spin relaxation time, $T_2$, and the magnetic field inhomogeneity due to defects and the magnet. The Curie $T$ dependence of the
NMR signal is described by the sample temperature, $T_s$. $\text{NF}_{\text{rel}}$ is the receiver noise factor relative to a receiver at 300 K.

Clearly, low sample temperature, low detector noise, and highly $^{13}$C enriched graphene are required for an NMR study. Sample temperature down to 1 K is customary in solid state NMR and $\text{NF}_{\text{rel}}=1/8$ was reported for cryo-probe NMR \cite{FirstCryoProbe}. We estimate from NMR data on graphitic carbon \cite{DresselhausAP2002,RuoffNMR} that the $\text{FWHM}=1/\pi T_2^*$ is 50 ppm at 14 T of fully $^{13}$C enriched graphene giving $T_2^*=10 \text{ }\mu\text{s}$. This estimate assumes either a single graphene sheet or a set of graphene layers oriented alike. These factors give an LOD for $^{13}$C graphene of $8\cdot 10^{12}$ which corresponds to a surface of $0.63 \text{ mm}^2$.
We think that synthesis of a fully $^{13}$C isotope enriched graphene with such an area
(not necessarily of a single piece) is within reach as fully $^{13}$C enriched graphite was recently synthesized \cite{RuoffNMR}. We expect
that a dedicated NMR microcoil setup, prepared by lithographic methods \cite{vanBentumJMR2007} would further decrease the LOD value and the required graphene sheet area.

In summary, we generalized the ca\-no\-ni\-cal theory of hyperfine interaction between nucleus and conduction electrons for graphene.
The orbital part of the HFI differs from its
usual form as it does not involve the angular momentum. We identified three distinct regimes in graphene based on the NMR measurables: Fermi- and Dirac-gas phases, and the extreme quantum limit. We argue that NMR on graphene is within realistic reach.

We acknowledge useful discussions with P. Thalmeier, A. V\'anyolos and T. Ma.
Work supported by the Hungarian Scientific Research Fund
under grants OTKA K72613 and F61733. F. S. acknowledges the Bolyai programme of the Hungarian Academy of Sciences.


\bibliographystyle{apsrev}
\bibliography{graph_nmr}

\end{document}